\documentclass[epj]{svjour}
\usepackage{graphicx}
\usepackage{color}
\begin{document}
\title{Comment on the paper Eur. Phys. J. A (2019) {\bf 55}:150} 
\author{C. Gonzalez-Boquera \inst{1}, M. Centelles \inst{1}, 
X. Vi\~nas \inst{1} \and L.M. Robledo \inst{2,3}}
%
%
\institute{ Departament de F\'isica Qu\`antica i Astrof\'isica
and Institut de Ci\`encies del Cosmos (ICCUB), 
Facultat de F\'isica, Universitat de Barcelona, Mart\'i i Franqu\`es 1, E-08028 Barcelona, Spain
\and
Departamento de F\'isica Te\'orica and CIAFF,
Universidad Aut\'onoma de Madrid,
E-28049 Madrid, Spain
\and 
Center for Computational Simulation,
Universidad Polit\'ecnica de Madrid, 
Campus de Montegancedo, Boadilla del Monte, E-28660 Madrid, Spain}
\date{Received: date / Revised version: date}
%
\abstract{
The conclusions of the study published as Eur. Phys. J. A (2019) {\bf 
55}:150 questioning the adequacy of the recently proposed Gogny D1M* 
interaction for finite-nuclei calculations using harmonic oscillator 
(HO) basis are revised. 
The existence of an instability in finite nuclei for D1M* 
when discretized coordinate-space methods are used to solve the HF equations (as 
shown in  Eur. Phys. J. A(2019){\bf 55}:150) is independently confirmed 
using a computer code based on a quasilocal approximation (QLA) to 
the HF energy density with finite-range forces. We confirm that the most 
affected quantity in the coordinate-space calculation is the spatial density 
at the origin. 
Our study reveals 
that some findings concerning these instabilities are not easy to 
reconcile with the arguments used in Eur. Phys. J. A (2019) {\bf 55}:150. 
For instance, some nuclei such as $^4$He and $^{40}$Ca, which diverge 
in HF mesh-point calculations performed with D1M*, become perfectly stable when 
Coulomb force is switched off. We have also found instabilities in some 
nuclei when the D1M interaction is used. Finally, a connection between 
the occupancy of $s$-orbitals near the Fermi level and the appearance 
of instabilities is observed. Several convergence and stability 
studies are performed with HO basis of different sizes and oscillator 
parameters to demonstrate the robustness of the D1M* results for finite 
nuclei when the HO basis is used.
\PACS{%
      { }{ }
     } 
} 
\maketitle

In Ref.~\cite{gonzalez18} we proposed a new parametrization of the 
Gogny interaction, denoted D1M*, aimed to obtain a stiffer equation of 
state (EoS) of neutron-star matter. The goal was to get maximum neutron 
star masses of $2 M_\odot$, in agreement with recent astrophysical 
observations \cite{demorest10,antoniadis13}. We were motivated by the fact that
this property is not achieved by any of the standard Gogny forces of the D1 family 
\cite{gonzalez17}. We also wanted to preserve the good description of 
the properties of spherical and deformed nuclei provided by the D1M force in 
Hartree-Fock-Bogoliubov (HFB) level \cite{goriely09}. 
In the fit of D1M* \cite{gonzalez18}, we modified the eight 
finite-range strength parameters of the D1M force while keeping the other parameters at their
nominal D1M values. Seven linear combinations of these strength parameters, related to 
different properties of symmetric nuclear matter, and the strength of the pairing interaction in the
$S$=0, $T$=1 channel were constrained to maintain the same values as in  D1M. 
The remaining combination was used to modify the slope of the
symmetry energy and, therefore, the stiffness of the neutron matter EoS and the prediction for the 
maximum neutron star mass. Finally, the strength $t_3$ of the density-dependent term of the Gogny interaction
was fine tuned to improve the quality of the computed binding energies.
All of the finite-nuclei calculations in \cite{gonzalez18} were carried out with the 
\mbox{HFBaxial} code \cite{robledo02} 
using an approximate second-order gradient method to solve the HFB 
equations in a HO basis including up to 19 major oscillator
shells and the oscillator lengths adapted to the characteristic $A^{1/6}$ length-scale 
dependence with mass number $A$. It is to be noted that all the HFB calculations of spherical and deformed nuclei
with Gogny interactions have always been performed in a HO basis since the seminal paper
of Decharg\'e and Gogny \cite{decharge80}, including the calculations leading to the 
determination of the values of the parameters of the interaction. In particular, 
this is the case of the D1M interaction \cite{goriely09},
to which we compare our results.    
 
In a recent paper  Eur. Phys. J. A (2019) {\bf 55}:150 \cite{martini19}  
and its preliminary version arXiv:1806.02080v1 \cite{martini18}, it is found that 
both the D1M* and D1N parametrizations of the Gogny force are affected by spurious finite-size instabilities 
in the $S$=0, $T$=1 channel. These instabilities are detected through a fully 
antisymmetrized RPA calculation of the nuclear matter response 
functions based on the continued fraction technique \cite{depace16}. 
This procedure has already been applied in \cite{depace16} to the search of instabilities 
in standard Gogny forces with or without tensor terms. In agreement with the results of 
previous analyses for Skyrme functionals \cite{hellemans13}, it was concluded that 
the key quantity to detect spurious finite-size instabilities is the critical density $\rho_c$. 
It is claimed that these instabilities develop unphysical results in some properties of the nuclei, as for example in the proton and 
neutron densities, if  $\rho_c  \simeq 1.2 \rho_0 \simeq 0.20$ fm$^{-3}$ for a momentum transfer of about 2.5 fm$^{-1}$. This 
critical density  may be reached in HF calculations of some nuclei, as for example $^{40}$Ca. 
The instabilities of D1M* and D1N were predicted in nuclear matter \cite{martini18,martini19,depace16} and 
their appearance in coordinate-space calculations of spherical finite nuclei was confirmed in \cite{martini18,martini19} 
using the FINRES$_4$ computer code \cite{bennaceur18}. As a consequence of finite-size instabilities,
the neutron and proton  density profiles of nuclei largely vary from one iteration to the
next in the iterative solution of the non-linear HFB equations, without reaching
convergence. 

In our paper arXiv:1807.10159v1 \cite{gonzalez18a} we commented on the results of arXiv:1806.02080v1 \cite{martini18}
and provided additional (and in our opinion very relevant) information about the possible impact of the finite-size instability 
of the D1M$^*$ force on calculations of observables like  binding energies,
neutron and proton radii and density profiles of finite nuclei using a HO basis. 
As discussed in detail in \cite{gonzalez18a}, we have found several important facts that, in our opinion, 
cannot be easily explained with the arguments used in  Refs.~\cite{martini18,martini19}.
However, our arXiv:1807.10159v1 paper \cite{gonzalez18a}---available well 
before the submission date of Ref. \cite{martini19} to {\it European Physical Journal A}---is
never mentioned in Ref.~\cite{martini19}  although it was early reported to the authors of \cite{martini18,martini19}.

\begin{figure}[t]
\centering
\includegraphics[width=0.9\columnwidth,clip=true]{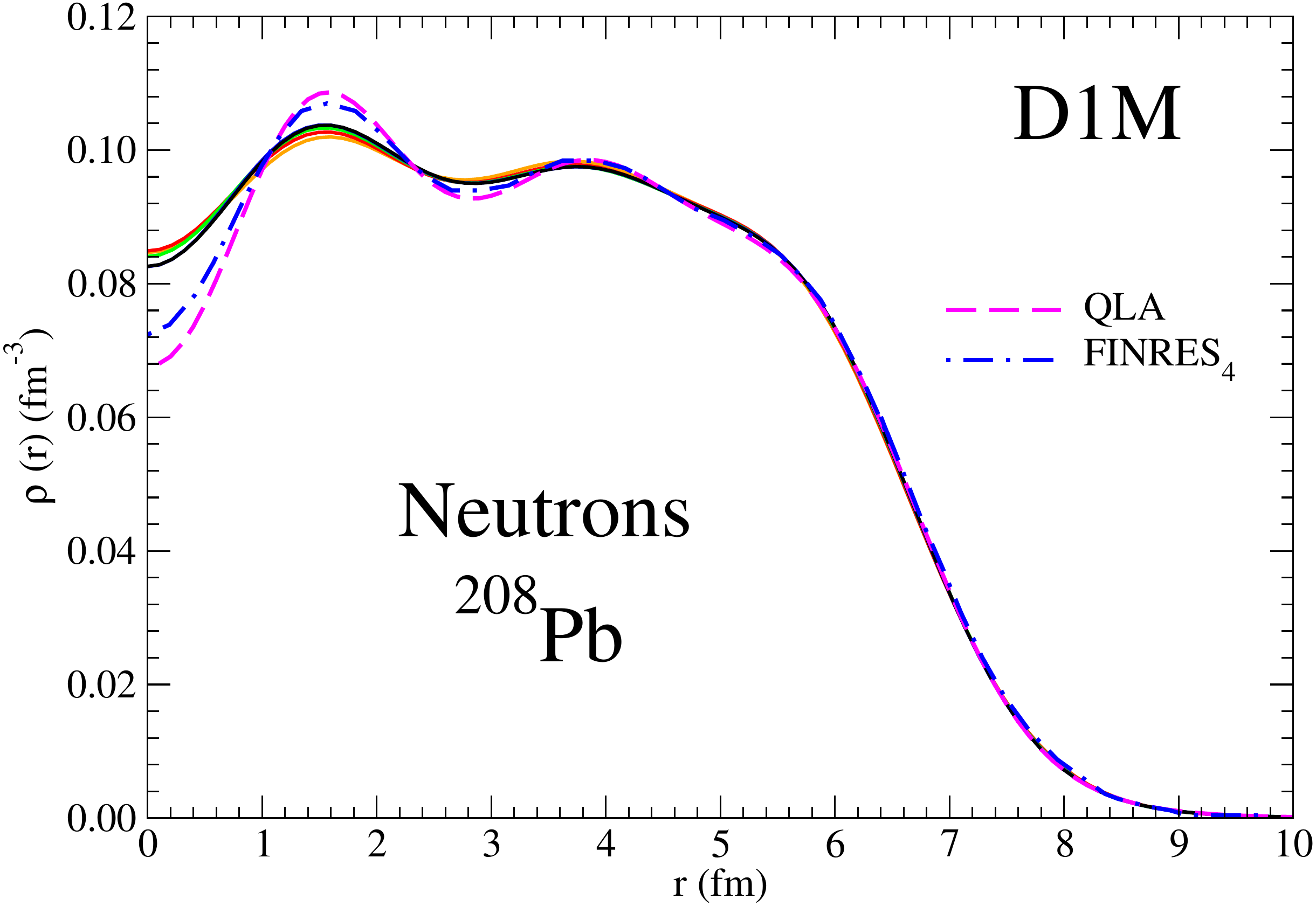}
\caption{Solid lines: Neutron density of $^{208}$Pb computed with the D1M force
with a HO basis with 12, 14, 16, 18 and 19 shells (yellow, red, green, blue and black curves, respectively). Dashed lines:
The same density obtained through a HF calculation on a mesh in the quasilocal
approach. Dash-dotted lines: The same density extracted from Fig.~3 in Ref.~\cite{martini18}
(HF calculation on a mesh with the FINRES$_4$ code \cite{bennaceur18}).}\label{fig:neutronesD1M}
\end{figure}
\begin{figure}[t]
\centering
\includegraphics[width=0.9\columnwidth,clip=true]{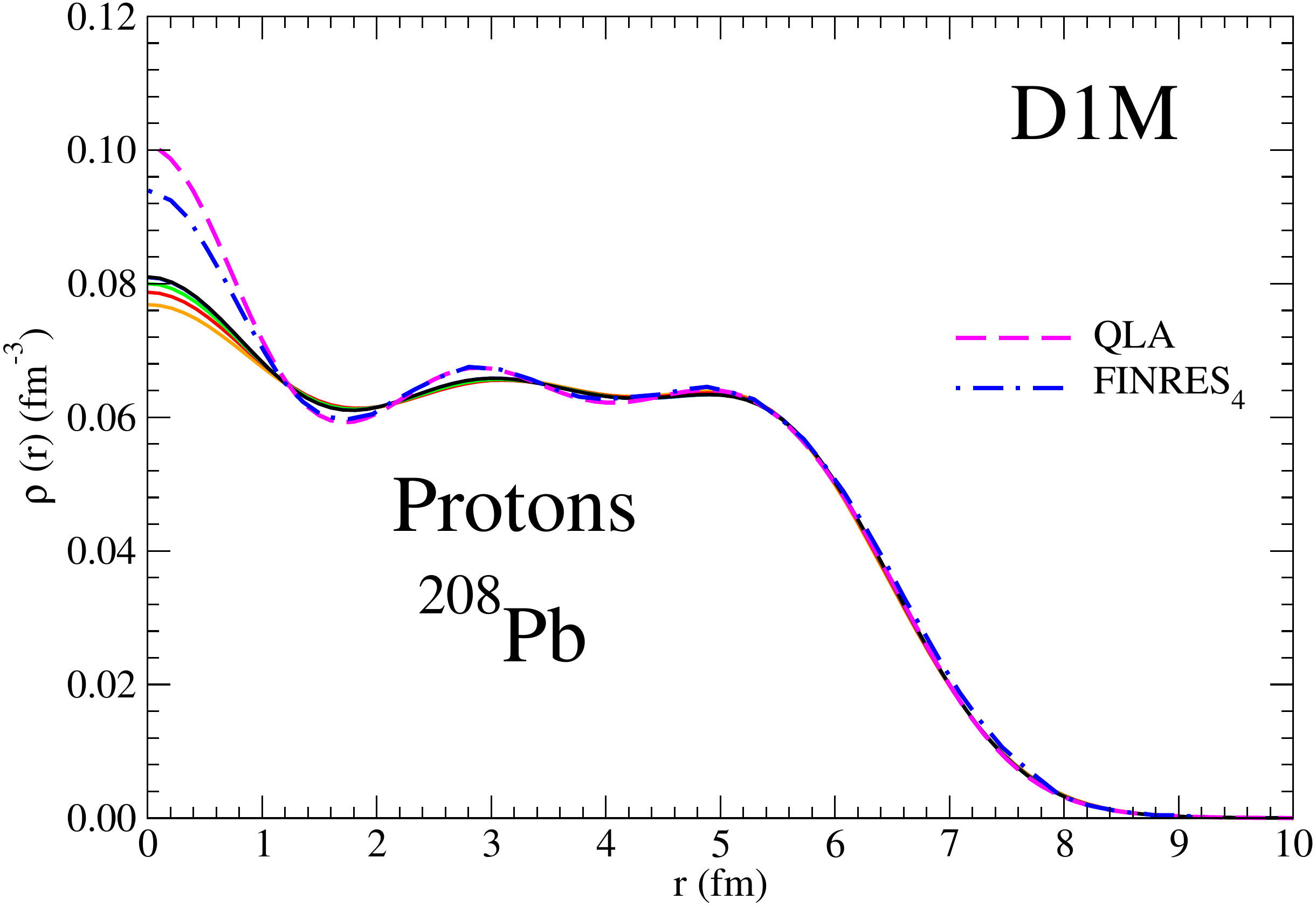}
\caption{The same as in Fig.~\ref{fig:neutronesD1M} but for the proton density.}\label{fig:protonesD1M}
\end{figure}
\begin{figure}[t]
\centering
\includegraphics[width=0.9\columnwidth,clip=true]{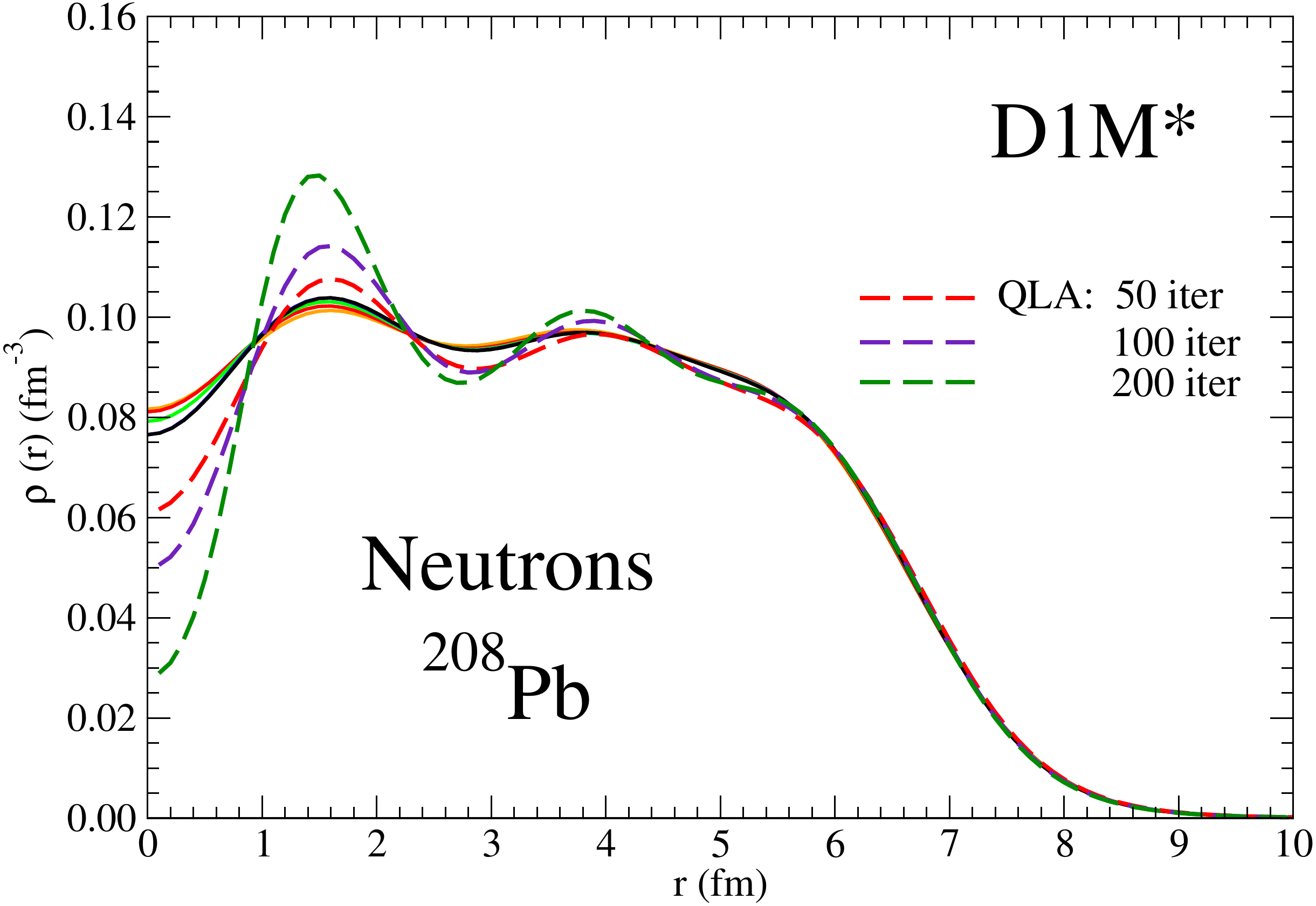}
\caption{The same as in Fig.~\ref{fig:neutronesD1M} but computed with the D1M* force.
The HF density is displayed for three different number of iterations.}\label{fig:neutronesD1Ms}
\end{figure}
\begin{figure}[t]
\centering
\includegraphics[width=0.9\columnwidth,clip=true]{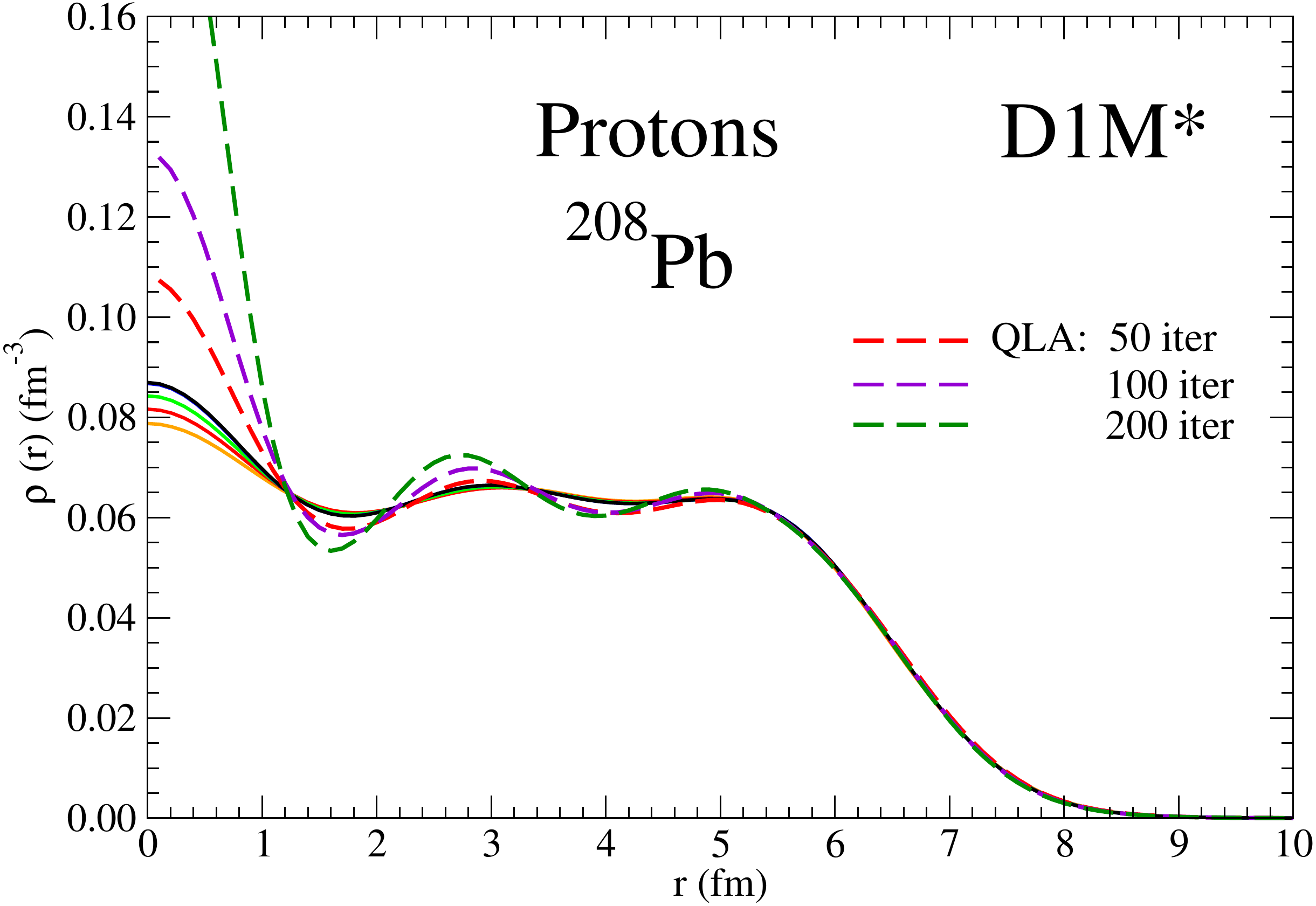}
\caption{The same as in Fig.~\ref{fig:neutronesD1Ms} but for the proton density.}\label{fig:protonesD1Ms}
\end{figure}

In order to independently confirm the results of \cite{martini18,martini19}
we have performed HF calculations with Gogny forces on a spatial mesh assuming spherical symmetry. To this end we use the QLA 
described in Ref.~\cite{soubbotin03}. In this approach  the HF exchange energy density is approximated by a quasilocal functional 
obtained using the extended Thomas-Fermi (ETF) density matrix \cite{soubbotin00}, which is 
similar to the  expansion for this quantity proposed by Negele and Vautherin \cite{negele72} or by Campi
and Bouyssy \cite{campi78}. In the QLA the energy density functional 
for finite-range effective interactions becomes local and therefore the HF equations in
coordinate-space take a similar form to those of Skyrme forces \cite{vautherin72}.
In Refs.~\cite{soubbotin03,krewald06,behera16} it has been shown that calculations in coordinate space using the QLA
provide results that are very close to the full HF values obtained with the HO basis. It is also important to point out that the HF 
results from  QLA accurately agree with those computed with the  
FINRES$_4$ code \cite{bennaceur18}. As an example, we show in Figs.~\ref{fig:neutronesD1M} and
\ref{fig:protonesD1M} the $^{208}$Pb neutron and proton density profiles calculated with the D1M interaction
using both coordinate-space codes, where the close agreement between the two methods is clearly seen. 
From Figs.~\ref{fig:neutronesD1M} and \ref{fig:protonesD1M} we also learn that the mesh calculations produce
density profiles with more pronounced oscillations near the center of the nucleus 
than the HO basis calculations (also shown in these figures for comparison).
This is a first qualitative indication that spherical densities calculated in a mesh may be more affected 
by the finite-size instabilities than the ones computed with a HO basis. We also find that mesh calculations with the QLA are 
well suited for the analysis of instabilities in spherical nuclei in coordinate-space calculations, as it reproduces all 
the instabilities of finite nuclei reported in \cite{martini18,martini19}. As an example, Figs.~\ref{fig:neutronesD1Ms} 
and \ref{fig:protonesD1Ms} show the proton and neutron densities of $^{208}$Pb calculated with the D1M$^*$ interaction and 
obtained after a given number of iterations using the QLA.
Clearly, in this case the mesh-point density profiles display a divergent behaviour with increasing number of iterations.

Let us now turn our attention to several conclusions that can be extracted from the mesh calculation that are barely or not 
discussed at all in \cite{martini18,martini19}.
First, we have performed HF calculations with the D1M* force for spherical even-even symmetric nuclei from $N=Z=2$ to $N=Z=126$
on a mesh without Coulomb interaction. To do these calculations we start from a trial symmetric Woods-Saxon potential and impose along the 
whole calculation  the same number of neutrons as protons, 
so that isovector contributions are completely eliminated.
These HF calculations are perfectly stable and the corresponding densities are 
fully converged after a large number of iterations ($\sim\,$10000 with a mixing factor 0.9  \cite{gonzalez18a}).
This seems to be in contradiction with, or unexplained by, the claim of \cite{martini18,martini19}
relating the instabilities to a critical density $\rho_c$ in 
nuclear matter. For instance, in the case of $^{4}$He and $^{40}$Ca including Coulomb effects, the HF results are unstable but turning 
the Coulomb force off makes the results completely stable in spite of the fact that the central proton and neutron densities are almost 
identical in the charged and uncharged cases. 
This fact corroborates that the origin of the instabilities 
is related to the isovector sector of the interaction.
On the other hand, it is to be noted that there exist nuclei with D1M*, 
such as $^{16}$O, $^{100}$Sn or the very asymmetric $^{176}$Sn, that fully converge in 
the coordinate-space calculation, even with the Coulomb interaction switched on. 
This has been verified with both our QLA code and the FINRES$_4$ code \cite{bennaceur18a}.
It tells us that the asymmetry of the nucleus is not a sufficient condition for developing the 
finite-size instabilities and that there can be other factors, such as the structure of the nucleus,
as we shall discuss below, that also play a role.

We have also detected that when the D1M interaction is applied in coordinate-space calculations
some nuclei such as $^{52}$Ca,  $^{54}$Ca,  $^{56}$Ca, $^{54}$Ti, $^{56}$Ti,  $^{58}$Fe,  $^{60}$Fe,  
$^{62}$Fe,  $^{60}$Ni and  $^{62}$Ni are unstable at the HF level. As the critical 
density $\rho_c$ for D1M is about 1.35$\rho_0$  \cite{martini18,martini19}, our finding is in 
contradiction with the criterion proposed in Refs.~\cite{martini18,martini19} because D1M should be stable according to~it.

\begin{table}[t]
\centering
\caption{For the D1M and D1M* Gogny interactions, Hartree-Fock binding energies 
obtained from the HO-basis calculation, 
the coordinate-space quasilocal calculation (QLA)
and the full coordinate-space calculation (FINRES$_4$) \cite{bennaceur18a}.
The percentage deviation of the HO-basis energy from the 
coordinate-space energy is shown in brackets. }
\begin{tabular}{rrrr}
\hline
\hline
  &\small $B_{\rm HO}$(MeV) &\small $B_{\rm QLA}$(MeV)~~ &\small $B_{{\rm FINRES}_4}$(MeV) \\
\hline
\hline\\[-3mm]
D1M        &          &            &          \\
$^{16}$O   & 128.02   &  127.02 (0.79\%)  & 128.07  (0.04\%)  \\
$^{132}$Sn & 1102.57  & 1103.31 (0.07\%)  & 1104.29 (0.16\%) \\
$^{208}$Pb & 1636.08  & 1637.96 (0.11\%)  & 1639.31 (0.20\%) \\
\hline\\[-3mm]
D1M*       &          &            &          \\
$^{16}$O   & 128.32   &  127.29 (0.81\%)  & 128.58  (0.21\%) \\         
$^{100}$Sn & 827.98   &  824.71 (0.40\%)  & 829.08  (0.13\%) \\
$^{176}$Sn & 1146.15  & 1146.26 (0.01\%)  & 1147.51 (0.12\%) \\
\hline
\end{tabular}
\label{table-ener}
\end{table}

In the upper part of Table \ref{table-ener} we report the binding energies of the nuclei  $^{16}$O, $^{132}$Sn and $^{208}$Pb
calculated with the D1M interaction using a HO basis \cite{robledo02}, the FINRES$_4$ code \cite{bennaceur18}
and the QLA used in this work. We can see that the HF binding energies computed with FINRES$_4$ 
are slightly larger than the ones provided by the HO basis, as can be expected. On the other hand,
the QLA results are in excellent agreement with those obtained in full HF calculations using the HO basis or the FINRES$_4$ code,
being the differences less than 1\% for all considered nuclei. A similar situation is found for the nuclei $^{16}$O, $^{100}$Sn 
and $^{176}$Sn computed with the D1M$^*$ force, where the corresponding binding energies are given in the lower part of 
Table \ref{table-ener}. These nuclei are found to be stable in coordinate space by independent calculations performed with the QLA and the 
FINRES$_4$ codes \cite{bennaceur18a}. The agreement between the HO basis and mesh results is again excellent in D1M* 
when the mesh calculations converge, which further supports the reliability of using the HO basis approach with D1M*. 
Regarding unstable nuclei in coordinate space, we have found, as discussed in \cite{gonzalez18a}, empirical evidence  
that the appearance of instabilities in finite nuclei is directly related to the presence of $s$-orbitals in the neighborhood of the 
Fermi level. This is, for example, the case in the nuclei $^4$He and $^{40}$Ca (neutrons and protons)
and $^{208}$Pb (protons) computed with D1M*.
However, the nuclei $^{16}$O, $^{100}$Sn or  $^{176}$Sn, for which the $s$-orbitals are far from the Fermi level,
are stable with the same D1M* force. 
A paradigmatic example is the case of the nuclei $^{22}$O and $^{24}$O. At HF level, the Fermi level of 
$^{22}$O is placed at the 1$d_{5/2}$ orbital and this nucleus is stable,  whereas 
the nucleus $^{24}$O has its Fermi level in the 2$s_{1/2}$ orbital 
and it is unstable. We have also considered the case where pairing correlations are taken into account through
a HF+BCS mesh-point calculation within the QLA \cite{krewald06}. Notice that this differs from the pairing calculations for open-shell nuclei 
of Ref.~\cite{martini19} performed at HFB level. The presence of excited $s_{1/2}$ energy levels in the single-particle spectrum  
considered for the BCS space makes the calculation unstable when the $s_{1/2}$ level is close to the Fermi level,
in spite of the fact that the calculation is stable at HF level. If the occupation of the $s_{1/2}$ level is small,
i.e., this level is far away from the chemical potential, the HF+BCS calculations may be stable.

Finally, let us point out that, in spite of the non-convergent behavior of the nucleon density profiles
when the mesh-point calculation is unstable,
there is a range of values in the number of iterations for which integrated quantities such as the
total binding energies present a smooth plateau pattern where the energy remains almost constant.
The quantities in the calculation  will ultimately oscillate when the number of iterations grows.
We find \cite{gonzalez18a} that the number of iterations for which the energy remains in the
plateau region weakly depends on the initialization conditions whereas it is strongly affected by
the mixing factor. This number also strongly depends on  the considered nucleus,
and, in particular, on the relative position of the $s_{1/2}$ levels with respect to the Fermi energy.
For a detailed analysis of the plateau pattern in the energy and other integrated quantities such as rms radii, or the 
virial theorem, we refer the reader to \cite{gonzalez18a}.

As established in Refs.~\cite{hellemans13,martini18,martini19}, RPA calculations 
of the nuclear matter response function allow one to detect
in a rather efficient way instabilities that prevent obtaining fully
converged self-consistent results in finite nuclei in coordinate space 
using some effective interactions, such as D1M* and D1N. This criterion may fail for some given 
interactions like D1M, which should be stable according to it. Therefore, our findings point out
that the problem is more involved and that additional investigations 
about the open questions suggested in this Comment are required. In this respect,
it would be worth to extend the analysis of Ref \cite{Pas15}, based on finding 
imaginary solutions of the RPA in finite nuclei, to the present case involving
finite range interactions.

\begin{figure}[t]
\centering
\includegraphics[width=0.9\columnwidth]{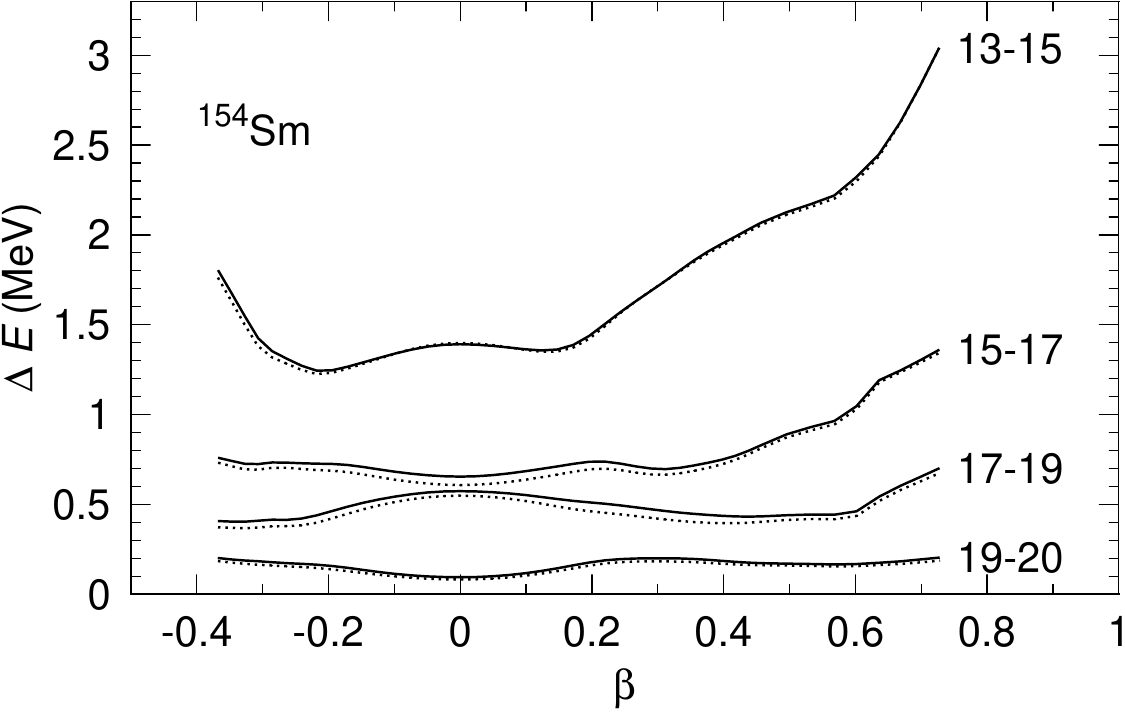}
\caption{Energy difference $\Delta E =E_\mathrm{HFB}(N)-E_\mathrm{HFB}(N')  $ between the HFB energies computed with different number 
of HO  shells and plotted as a function of the
quadrupole deformation parameter $\beta_{2}$ for the nucleus $^{154}$Sm.
The number of HO shells $N$ and $N'$ for each curve is indicated as labels.
Full (dotted) lines correspond to the D1M (D1M*) parametrizations.}\label{fig:154Sm}
\end{figure}

In the paper where D1M* was proposed \cite{gonzalez18}, we did 
finite-nuclei HFB calculations in a HO basis with the HFBaxial code, to fine-tune the 
density-dependent strength as to improve the agreement of
binding energies with experimental data. 
For those calculations we used a basis with a number of shells 
depending on the region of the nuclear chart. The calculation covered both deformed and spherical 
nuclei and we computed the properties of more than 600 even-even nuclei. In this calculations
we did not observe any convergence issue after over 30000 HFB calculations. The stability of the HO basis
calculations could be related to the ultraviolet cutoff in momentum space inherent to the HO basis.
The ultraviolet cutoff  serves as a regulator 
for problems related to high-momentum components in the wave
function in a way that closely resembles other regulators in pairing
calculations with zero range forces---as, for instance, the introduction of an active window. 
On the other hand, mesh calculations, with the use of finite
differences to evaluate derivatives, are  prone to suffer the effect of
those ultraviolet problems. 
This difference between HO basis and mesh calculations 
was already recognized in \cite{martini18,martini19}, where it was argued that the use of a 
HO basis ``strongly renormalizes the interaction and inhibits
the development of instabilities". However, the authors of \cite{martini18,martini19}
conclude that ``the D1M* interaction should only be used with the basis 
employed to fit its parameters", a statement that, according to our 
experience, only applies to the binding energy of the nucleus (the variational quantity) and not
to the rest of the observables. 
From the statement of \cite{martini18,martini19} one should expect 
significant changes in the value of physical observables computed with 
D1M* as the HO basis size is changed. However, this is not the
case for typical HO basis: we have carried out calculations 
including 11, 13, 15, 17, 19 and 20 full HO shells for some
representative nuclei using both D1M* and D1M. The range of nuclei considered
includes deformed nuclei like $^{224}$Ra, $^{168}$Er or $^{48}$Cr 
and spherical nuclei like $^{16}$O, $^{40}$Ca, $^{56}$Ni, $^{100}$Sn, $^{132}$Sn 
or $^{208}$Pb. Except for the binding energy (which is the variational 
magnitude and therefore always increases with increasing basis size), 
the changes in the other observables
(radii, quadrupole deformation, octupole deformation, excitation energy of the lowest
quasiparticle, etc.) are of the order of a few in a thousand when going 
from the smallest to the largest basis. Interestingly, the convergence rate with basis size of the
density at the origin is rather slow and requires a large number of shells
both in D1M and D1M*, and even in the D1S case, as can be seen in 
Figs.~\ref{fig:neutronesD1M}--\ref{fig:protonesD1Ms}. It is to be pointed out that the central density does not 
enter significantly in most of the observables like radii or multipole
moments as the corresponding operators go to zero at the origin. Also 
the energy, which should be more sensitive through the strongly repulsive
density-dependent part of the interaction to the slow convergence rate of the central density, shows
a smooth behavior. On the other hand, in Ref.~\cite{PRC89.054310} we
studied fission properties of the uranium isotopes including very neutron-rich 
isotopes using, among others, the parametrization D1N, which is known to show instabilities~\cite{martini18}.
In those calculations we used a HO basis very different from the one used
in the D1N fit to ground-state properties and never observed any significant 
deviation in the shape and properties of the potential energy surfaces from
the ones obtained with the D1S and D1M parametrizations. As additional
evidence, we show in Fig.~\ref{fig:154Sm} the difference in HFB energies 
$\Delta E$ when obtained with different number of HO shells 
($\Delta E=E_\mathrm{HFB}(N)-E_\mathrm{HFB}(N'))$. The results are obtained 
and plotted as a function of the quadrupole deformation parameter 
$\beta_{2}$ for the nucleus $^{154}$Sm. To simplify the discussion the same
oscillator lengths are used in the whole deformation interval and therefore
the convergence of the relative energies is slower than in standard calculations.
The two sets of curves correspond to D1M (full) and D1M* (dotted) and we
observe they almost coincide in all the cases. All the results
presented above constitute strong evidence that there is an ample range
of valid HO basis where the results are converged and consistent.

Our approach is rather pragmatic as there is no formal justification of its validity in
a regularization-renormalization framework. It would be very illuminating to
study the UV regularization inherent to the HO from this perspective, but we
feel such an study lies well beyond the scope of the present comment.

At this point it is worth to mention a difficulty of the HO basis  connected
with the evaluation of matrix elements of a two-body Gaussian interaction for large
values of the HO quantum numbers.
The standard expressions \cite{Girod,Younes,Egido} for those matrix elements are given in terms
of finite alternating-in-sign sums. For large values of the HO quantum numbers the
terms in the sums become very large and the alternating sign
leads to unwanted loss of accuracy \cite{Egido} 
\footnote{To understand the 
problem, let us imagine a calculation carried out with 64-bit floating 
point arithmetic with 13 digits accuracy. If the alternating sign sum
involves terms which are 13 orders of magnitude larger that the result of
the sum, then the numerical error is of the order of the sum.}. 
This effect starts to be relevant for 22-24 HO shells
and can easily turn  a repulsive matrix element into an attractive one. 
This is not a limiting problem for the HO basis as  calculations with the Gogny force are
in most of the applications well converged already with 22 shells \footnote{It is possible to 
reach 26 shells depending on the nucleus and the oscillator lengths---a typical example is fission
where 26 shells are used in the $z$ direction but with a large oscillator length).}. 

\begin{figure}[t]
\centering
\includegraphics[width=0.85\columnwidth]{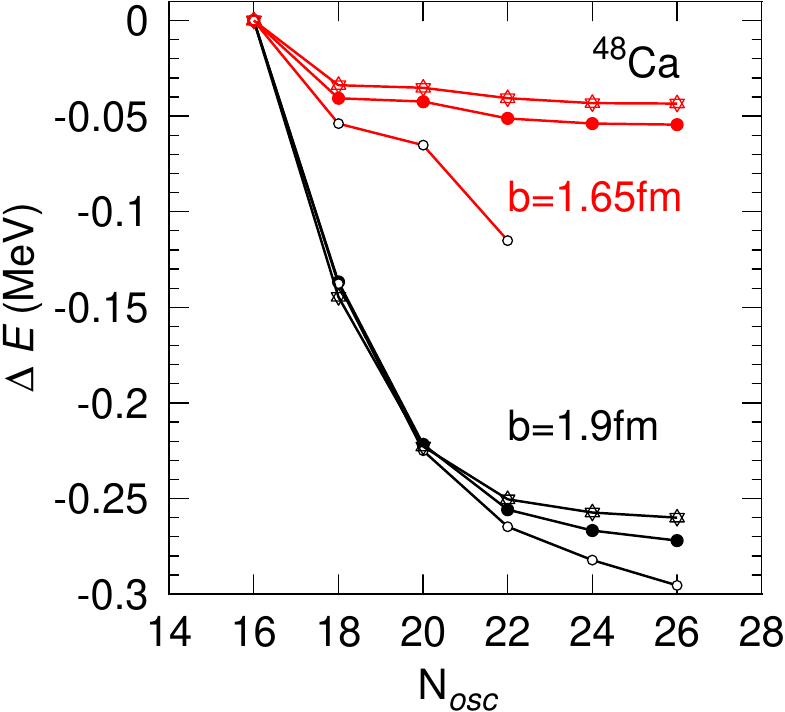}
\caption{For the nucleus $^{48}$Ca we show the difference in the total energy with 
respect to the 16 shell calculation as a function of the number of shells. Results
for D1S (stars), D1M (bullets), and D1M* (open circles) are shown for two
different sets of oscillator lengths ($b=1.9$~fm, black curves; $b=1.65$~fm, red curves).
Note the small range of the vertical scale compared with the total energy of $^{48}$Ca.}\label{fig:48Ca}
\end{figure}

Taking the previous considerations into account, it is now possible to understand the results of
Fig.~\ref{fig:48Ca} where the energy difference with
respect to the reference 16 shell calculation is shown as a function of the number of shells
for the nucleus $^{48}$Ca. This figure is similar to Fig. 3 of \cite{martini19}. 
We show results for two oscillator lengths, one is $b=1.65$~fm and corresponds to the minimum of the HFB energy with
16 shells (red curves). The other corresponds to $b=1.9$~fm. In this case
the reference HFB energy at 16 shells is higher than the one for $b=1.65$~fm. 
This is the reason why, in the
plot, the two set of curves do not converge at the same value of the binding energy
with 26 shells. We have tested that with 26 shells the binding energies with
different oscillator lengths coincide at the level of a few keV.
For $b=1.65$~fm we observe a peculiar behaviour in the three parametrizations  D1S, D1M and D1M*
at $N_\textrm{osc}=20$ 
that could be a consequence of the inaccurate  evaluation of 
matrix elements. At $N_\textrm{osc}=22$ D1M* shows a dip and at $N_\textrm{osc}=24$
the HFB calculation does not converge leading to wild numbers. This lack of 
convergence should not be attributed to finite size instabilities as it is
suggested in \cite{martini19} but rather to the inaccurate 
evaluation of  matrix
elements for large HO quantum numbers. 
The $b=1.9$~fm calculations seem much more
stable and show in the three cases a good convergence rate with $N_\textrm{osc}$. 
In D1M* the convergence rate seems to be slower than for D1S and D1M.
From the above results, it is clear that a stable and
consistent solution to the problem of evaluating matrix elements of a 
finite-range Gaussian interaction for large oscillator quantum numbers is required.

Let us also mention that a study of the convergence of calculations
with the number of shells with Skyrme forces was carried out in \cite{hellemans13}.
In an spherical calculation and for contact forces they were able to reach
60 shells. As the HO basis is complete, in the limit of infinite number of shells
the HO results should be equivalent to the ones on a mesh 
and therefore the appearance of difficulties with 50 or 60 shells is not surprising. For such
a large number of shells the ultraviolet cutoff is increased and the regularization
property of the HO basis is weakened. Comparing with the pairing case, if we
increase the active window size the results will be unphysical. There is another
difference with the present case: the expressions for the matrix elements
of contact interactions in a HO basis differ from the ones obtained for Gaussian interactions 
and seem to be less likely to suffer from the numerical instabilities discussed
above.

To finish the Comment let us summarize our main findings
\begin{itemize}
\item The QLA represents an alternative approximate method to signal the existence
of instabilities in discretized space coordinate calculations. It is much faster
than the exact solution because of the local treatment of the exchange term.
\item Our results independently confirm the results of \cite{martini19} pointing
to a finite-size instability in discretized coordinate-space calculations in D1M*,
D1N and D1M (see below).
\item There are several examples of nuclei which are unstable when computed with
the D1M interaction. As the critical density $\rho_c$ of D1M is $1.35\rho_0$ this
interaction should remain stable with the criterion of Refs.~\cite{martini18,martini19}.
\item When the discretized coordinate-space calculation converges the results 
obtained with FINRES$_4$, QLA and a HO basis are consistent.
\item The appearance of instabilities seems to be connected with the presence
and occupancy of $s$ orbitals near the Fermi level.
\item The HO basis with its UV cutoff acts as a regulator of the instability
and provides consistent results compatible with the ones obtained with D1S. 
There is an ample range of HO basis sizes that can be used in the calculations
to accommodate different deformation regimes that produce essentially the same
values of non-variational observables.

\end{itemize}
We would also like to mention some supplemental results that seem to reinforce
our conclusions
\begin{itemize}
\item Even-even $N=Z$ nuclei turn to be stable if Coulomb is switched off. In most of
the calculations the value of the central density is similar to the one with Coulomb.
Therefore, these calculations suggest that the value of the 
central density should not play a central role in determining the existence of instabilities. 
\item There are examples of nuclei for which the calculation is stable with very
different values of $N-Z$.
\item There exist a plateau with the number of iterations of the selfconsistent 
calculation where physical quantities remain reasonably constant. The width of the
plateau depends on the nucleus, contrary to the naive expectation that the instability 
should develop in the very early stages of the iterative process. 
\item A problem found in \cite{martini19} in a HO basis calculation is attributed
to a numerical problem rather than an instability of the interaction.
\end{itemize}

We conclude that the criterion proposed in \cite{martini19} is not enough to 
signal the existence of instabilities (D1M case) and there are other finite
nuclei effects like the position of the $s$ orbitals relative to the Fermi 
level that are relevant. The HO basis acts as a UV regulator that allows 
consistent calculations of finite nuclei with D1M* and all the other
interactions like D1N or D1M showing instabilities.
Such calculations will serve to elucidate the role played by the slope of the symmetry energy in 
determining nuclear structure properties.

\section*{Acknowledgments}

The authors are very grateful to K. Bennaceur for exchange of useful information.
C.G., M.C., and X.V. were partially supported by Grant FIS2017-87534-P from MINECO and FEDER
and Project MDM-2014-0369 of ICCUB (Unidad de Excelencia Mar\'{\i}a de Maeztu) from MINECO.
C.G. also acknowledges Grant BES-2015-074210 from MINECO.
The work of LMR was partly supported by the Spanish
MINECO Grant No.~FPA2015-65929, No.~FIS2015-63770 and No. PGC2018-094583-B-I00.

\end{document}